# Unconventional anomalous Hall effect in epitaxially stabilized orthorhombic Ru$^{3+}$ perovskite thin films

L.-F. Zhang,[1] T. C. Fujita,[1,*] Y. Masutake,[2] M. Kawamura,[3] T. Arima,[3,4] H. Kumigashira,[2,5] M. Tokunaga,[3,6] and M. Kawasaki[1,3]

[1]Department of Applied Physics and Quantum Phase Electronics Center, University of Tokyo, Tokyo 113-8656, Japan.

[2]Institute of Multidisciplinary Research for Advanced Materials (IMRAM), Tohoku University, Sendai 980-8577, Japan.

[3]RIKEN Center for Emergent Matter Science (CEMS), Wako 351-0198, Japan.

[4]Department of Advanced Materials Science, University of Tokyo, Kashiwa 277-8561, Japan.

[5]Photon Factory, Institute of Materials Structure Science, High Energy Accelerator Research Organization (KEK), Tsukuba 305–0801, Japan.

[6]The Institute for Solid State Physics, The University of Tokyo, Kashiwa 277-8581, Japan.

*corresponding author

**Abstract**

Complex oxides are mesmerizing material systems to realize multiple physical properties and functionalities by integrating different elements in a single compound. However, owing to the chemical instability, not all the combinations of elements can be materialized despite the intriguing potential expected from their magnetic and electronic properties. In this study, we demonstrate an epitaxial stabilization of orthorhombic Ru$^{3+}$ perovskite oxides： LaRuO$_3$ and NdRuO$_3$, and their magnetotransport properties that reflect the difference between non-magnetic La$^{3+}$ and magnetic Nd$^{3+}$. Above all, an unconventional anomalous Hall effect accompanied by an inflection point in magnetoresistance is observed around 1.3 T below 1 K for NdRuO$_3$, which is ascribed to topological Hall effect possibly due to a non-coplanar spin texture on Nd$^{3+}$ sublattice. These studies not only serve a new testbed for the interplay between spin-orbit coupling and Coulomb interaction but also open a new avenue to explore topological emergent phenomena in well-studied perovskite oxides.

**Introduction**

The interplay between spin-orbit coupling (SOC) and Coulomb correlation has become a central topic in condensed-matter physics (*1–3*). This crucial importance has been widely known, especially in 4*d* and 5*d* transition metal oxides, where SOC acts on an energy scale comparable to the other energy scales such as bandwidth, crystal field, and Coulomb interaction, etc. This class of compound system has attracted considerable attention since the discovery of Mott insulating state in $Sr_2IrO_4$ (*4*). The study has expanded to other heavy transition metal oxides, especially with the $d^5$ configuration under octahedral crystal field (*1, 2, 5*). In such materials of interest, partially filled 4*d* or 5*d* $t_{2g}$-band are split into an effective $j = 1/2$ doublet and effective $j = 3/2$ quartets as shown in Fig. 1A. Indeed, a number of novel topological phenomena related to the effective $j = 1/2$ states have been reported not only in oxides with $Ir^{4+}$ (*6–10*) but also in α-$RuCl_3$ with $Ru^{3+}$ as a possible candidate of Kitaev quantum spin liquid (*11–14*). Here, we pose a question whether it is possible to stabilize $Ru^{3+}$ state in oxide thin films as in the case of well-studied $Ir^{4+}$, providing more "knobs" to tune the physical properties by utilizing epitaxial strain or heterointerface.

From the chemistry point of view, the stable oxidation state of Ru is generally +4 or higher in oxides (*15*). *Ln*$RuO_3$ (*Ln*: lanthanide) perovskite oxides (orthorhombic *Pbnm* phase, as shown in Fig. 1B) are rare exceptions of an oxidation state of $Ru^{3+}$. Because of difficulties in synthesis, however, studies on *Ln*$RuO_3$ are quite limited; with bulk polycrystals (*16–19*) and single crystals (*20*), and their electronic and magnetic properties have not been revealed well, in contrast to an immense amount of previous literature on *Ae*$Ru^{4+}O_3$ (*Ae*: alkaline earth) (*21–25*). Different from *Ae*$Ru^{4+}O_3$, *Ln*$RuO_3$ can accommodate magnetic $Ln^{3+}$, which renders this system an intriguing playground for the magnetic interaction between Ru-4*d* itinerant electrons and *Ln*-4*f* localized moments (Fig. 1C). In this point of view, *Ln*$RuO_3$ is a unique system among *Ln*-*M*-O (*M*: transition metal) complex oxides that can support magnetization from *Ln*-site and metallic conduction on *M*-O network at the same time, which is critical for evaluating the magnetic interaction by magnetotransport properties. To address this issue, we summarize electrical conductivity and crystal phase in terms of two well-studied structures; perovskite (*LnMO*$_3$) and pyrochlore (*Ln*$_2M_2O_7$) in Fig. 1D. Most of *LnMO*$_3$ and *Ln*$_2M_2O_7$ compounds are insulating; metallic ones are limited to $LaNiO_3$, $LaCuO_3$, LRO, $Pr_2Ir_2O_7$, and $Ln_2Mo_2O_7$ (*Ln* = Nd, Sm, Gd) (*16, 26–29*). In this regard, we are motivated to explore *Ln*$RuO_3$ to provide a new tunable platform for emergent magnetotransport phenomena originating from correlated electrons with both sizable SOC and magnetic interaction.

In this work, we report the fabrication of the epitaxial thin films of LRO and $NdRuO_3$ (NRO) by making use of a solid phase epitaxy technique. X-ray absorption spectroscopic (XAS) studies supports $Ru^{3+}$ oxidation state in the films. We carry out systematic electrical transport studies in magnetic fields up to 54 T and at temperatures down to 50 mK to investigate the veiled magnetotransport properties of LRO and NRO. We find that both LRO and NRO are metallic though NRO shows a slight upturn below 20 K. LRO exhibits only an ordinary Hall effect, while NRO exhibits a clear signature of anomalous Hall effect (AHE), which is successfully deconvoluted from ordinary one with the aid of the high-filed measurements up to 54 T. Especially below 1 K, an unconventional AHE emerges at around 1.3 T concomitantly with an anomaly in magnetoresistance. We propose that this unconventional AHE is a topological Hall effect due to a non-coplanar spin texture and resultant finite scalar spin chirality realized in the orthorhombic perovskite structure. These results altogether demonstrate that epitaxial stabilization is a potent activator to push new frontiers of the playground for strongly correlated electrons (*30–32*) that will facilitate new fundaments and high-end functional applications (*33–35*).

**Results**

**Structural properties**

Epitaxial LRO and NRO thin films were prepared on STO (001) substrates by pulsed laser deposition (PLD) and subsequent annealing process (See Materials and Methods and Figs. S1 and S2 for the details). XRD $2\theta$-$\theta$ scan profiles of the LRO and NRO thin films on SrTiO$_3$ (STO) (001) substrates are presented in Figs. 2A and 2B, where the peaks of thin films are indicated by triangles and squares, respectively. From the positions of (002) peaks (pseudocubic setting), out-of-plane lattice constants of LRO and NRO are deduced to be 3.96 Å and 3.94 Å, respectively. The full width at half maximum (FWHM) in rocking curves of (002) peaks (not shown) is less than 0.1° for both LRO and NRO films, reflecting high orientation and crystallinity. Thicknesses of the LRO and NRO films are around 6 nm and 9 nm, respectively, which are deduced from x-ray reflectivity measurements at lower incident angles (Fig. S1). The epitaxial relationship between the substrate and the thin films is clarified by the reciprocal space mappings (RSM) presented in Fig. 2C (and Fig. S4 in Supplementary Materials). The peak of LRO exhibits a slight broadening toward smaller $Q_x$ from the peak position of STO. Since the mismatch between bulk LRO (3.94 Å in pseudocubic setting) and STO substrate (3.905 Å) is as large as 1%, this broad peak indicates that the thin film near the interface is fully strained while the rest part is partially relaxed. On the other hand, NRO (3.92 Å in pseudocubic setting), which has less mismatch to STO substrate, does not show such a peak broadening and thus is fully strained. The high quality of the films can be confirmed by TEM images presented in Figs. 2D and 2E as well. Lattice images in an atomic resolution are clearly seen for thin films epitaxially grown on STO substrate. The formation of orthorhombic perovskite structure in LRO and NRO is also supported by electron beam diffraction (Fig. S3 in Supplementary Materials).

**Oxidation state of Ru**

As evidenced by the XRD and TEM measurements, it is apparent that LRO and NRO films are stabilized as perovskite structures. However, as mentioned above, $Ru^{3+}$ is generally unstable in oxides (*15*). Thus, to further confirm the oxidation state of Ru, XAS measurements are performed for both LRO and NRO films. Fig. 2F and its inset respectively show the wide and near-edge XAS for these thin films in comparison with those for Ru metal ($Ru^0$), $SrRuO_3$ ($Ru^{4+}$), and $RuO_2$ ($Ru^{4+}$) as references. As shown in the inset of Fig. 2F, Ru K-edge energies of LRO and NRO locate at the lower energy (reduction) side than those of $SrRuO_3$ and $RuO_2$, indicating that oxidation state of Ru in LRO and NRO is less than +4. The oxidation state is semi-quantitatively evaluated by comparing the Ru K-edge energies of the films with those of reference compounds for $Ru^{2+}$ or $Ru^{3+}$, as shown in Fig. 2G (*36*, *37*). Here, Ru K-edge energy is defined as the energy at which the normalized absorption intensity decreased by half. The Ru K-edge energies of LRO and NRO are represented by red and blue vertical lines, respectively, while the ones of other compounds are plotted by symbols with the reported oxidation states. One can see that the Ru K-edge energies of LRO and NRO are located among the ones of standard references for $Ru^{3+}$; $Ru^{3+}Cl_3$, $Ru^{3+}I_3$, and $[Ru^{3+}(NH_3)_5]Cl_3$. From this result, we conclude that Ru predominantly exists as +3 oxidation states in both LRO and NRO films.

**Comparison of magnetotransport properties between LaRuO₃ and NdRuO₃**

Temperature dependence of longitudinal resistivity ($\rho_{xx}$–$T$ curves) for LRO and NRO is presented in Figs. 3A and 3B. In Fig. 3A, $\rho_{xx}$ is shown in a logarithmic scale to compare with those in previous studies (dashed lines). For LRO, our thin film sample is metallic down to 5 K and $\rho_{xx}$ slightly increases at lower temperatures. Compared with the works by Bouchare *et al.* and Sugiyama *et al.* (*16*, *17*) for polycrystalline samples, $\rho_{xx}$ in our LRO film has a lower resistivity by one order of magnitude. For NRO, in striking contrast to the insulating behavior reported in the bulk polycrystalline sample (*18*), our epitaxial thin film is metallic down to 20 K and shows a weak upturn in $\rho_{xx}$ with further cooling. This remarkable improvement in conductivity, plausibly originating from the high crystallinity of our thin films, endows an opportunity to examine magnetotransport properties of the two contrasting LRO and NRO with non-magnetic $La^{3+}$ and magnetic $Nd^{3+}$, respectively.

Magnetic field ($B$) dependence of magnetoresistance ratio (MRR (%) ≡ $[\rho_{xx}(B)/\rho_{xx}(0) - 1] \times 100$) and Hall resistivity ($\rho_{yx}$) up to 9 T measured by Physical Properties Measurement System (PPMS, Quantum Design Co.) are presented in Figs. 3C–3F. For LRO, MRR increases monotonically (Fig. 3C), and $\rho_{yx}$ is linear (Fig. 3D) to the magnetic field down to 2 K. These behaviors suggest the paramagnetic and metallic state in LRO, which is consistent with previous reports (*16*, *17*). On the other hand, NRO exhibits peculiar behaviors at low temperatures in both MRR (Fig. 3E) and $\rho_{yx}$ (Fig. 3F). Negative MRR, which is a typical response in magnetic materials, is observed below 5 K. Above 30 K, $\rho_{yx}$ is linear but exhibits a sign reversal at around 60 K. Below 30 K, $\rho_{yx}$ is no longer linear to the magnetic field. What is more, at 0.5 K, a unique bulge in MRR and a hump structure in $\rho_{yx}$ are observed at an identical magnetic field of ~1.3 T (Also discussed in Figs. 5A and 5B). These distinct behaviors observed only in NRO at low temperatures suggest an essential role of an interaction between Nd-4$f$ spins and Ru-4$d$ electrons in magnetotransport properties, compelling us to perform further measurements with a help of a pulsed high magnetic field.

**Magnetotransport properties of NdRuO$_3$ under high magnetic field**

Magnetic field dependence of MRR and $\rho_{yx}$ up to 54 T for NRO are presented in Figs. 4A and 4B, respectively, together with the results measured with a PPMS up to 9 T (thick lines). As presented in Fig. 4A, MRR increases monotonically up to 54 T above 10 K. Below 10 K, MRR decreases at lower fields with a minimum at $B \approx 10$ T and turns to increase at higher fields similar to the ones above 10 K. This can be seen more clearly in Fig. 4D, where $\rho_{xx}$–$T$ curves under different magnetic fields are shown. Intriguingly, the slopes of $\rho_{yx}$ turn to saturate at a similar negative value at higher fields regardless of the measurement temperatures (Fig. 4B). Because of this, ordinary Hall coefficient ($R_H$) can be deduced from linear fittings of $\rho_{yx}$ between 45 T and 54 T. As presented in Fig. 4E, $R_H$ has little temperature dependency below 150 K. Therefore, with the help of the high-field measurements, the sign reversal of $\rho_{yx}$ observed in Fig. 3F is revealed to be irrelevant to a carrier-type change.

**Anomalous Hall effect of NdRuO$_3$**

Having clarified that the carrier-type of NRO is electron regardless of temperatures, we can discuss the AHE of NRO. With using $R_H$ deduced from the high-field measurements in Fig. 4E, anomalous Hall resistivity $\rho_{AHE}$ is defined by $\rho_{AHE}(B) = \rho_{yx}(B) - R_H B$ as shown in Fig. 4C. AHE is an important electrical transport phenomenon attracting extensive interest in both fundamental physics and potential applications (*38*, *39*). Indeed, NRO unveils various AHE originating from different magnetic interactions depending on temperatures and magnetic fields as we discuss below.

At high temperatures, AHE emerges below ~100 K while $\rho_{yx}$ is linear to the magnetic field above 120 K. Since 100 K is too high for $Ln^{3+}$ ions to be ordered in perovskite oxides (*40*), AHE in this temperature range is supposed to originate from the magnetism of Ru$^{3+}$ induced by the applied magnetic field. At lower temperatures, AHE at the high-field region shows almost no temperature dependence while it develops much faster at the low field region below 20 K. This feature can be also confirmed clearly in the temperature dependence of $\rho_{AHE}$ at various magnetic fields ($\rho_{AHE}(B)-T$ curve) shown in Fig. 4F. Here, as one can see clearly in the $\rho_{AHE}-T$ curve at 1 T, $\rho_{AHE}$ increases dramatically below 20 K, which is distinct from other curves measured at higher magnetic fields.

Finally, below 1 K, AHE starts to show a distinct behavior. Figure 5A shows $\rho_{AHE}$ below 5 K and 4 T. At 0.8 K and 0.5 K, clear hump structures are observed at ~1.3 T while $\rho_{AHE}$ almost monotonically increases as a function of $B$ above 1.5 K. Furthermore, an anomaly appears in MRR as well. As presented in Fig. 5B, inflection points are observed for MRR at the same field range below 0.8 K, which is absent at higher temperatures. To compare the relationship between these two anomalies, we show the magnetic field derivatives of $\rho_{AHE}$ and MRR at 0.5 K in Fig. 5C. The hump structure in $\rho_{AHE}$ and the inflection point in MRR appear almost at the same magnetic field, which is indicative of the same origin for these two phenomena: plausibly a magnetic one.

Since AHE is generally proportional to the magnetization $M$, it is didactic to compare the observed AHE with the magnetic properties measured for another thick NRO film (NRO5) down to 2 K (See Supplementary Text 5 and Fig. S7). As presented in Fig. S7C, $M$ becomes detectable at ~100 K, which coincides with the emergence of the AHE shown in Fig. 4C. In the range of 30–100 K, $M$ increases linearly to the magnetic field up to 7 T, indicating that NRO is in an AFM or a paramagnetic (PM) phase. Considering that there is no anomaly in the $\rho_{xx}-T$ curve usually concomitant with magnetic transitions, we speculate that NRO is PM in this temperature range, and magnetic moments of $Ru^{3+}$ induced by the applied magnetic field lead to the AHE. At lower temperatures than 20 K, $M$ becomes non-linear to the magnetic field, which can be the reason for the steep increase in AHE at the low-field region in Figs. 4C and 4F. Indeed, at $B = 1$ T, the temperature dependence of magnetization ($M-T$ curve) is scaled well with the $\rho_{AHE}-T$ curve for this thick NRO film as presented in Fig. S7E. This alludes to the existence of an additional magnetic component that is sensitive to magnetic field only at lower temperatures, and thus we speculate that it is from $Nd^{3+}$ moments, yet they may not be ordered in this measurement temperature range.

**Possible origin of unconventional anomalous Hall effect of NdRuO$_3$**

For the origin of the hump structure in AHE at lower temperatures than 1 K, one possible scenario is to consider a topological spin texture. Apparently, this anomaly cannot be explained in terms of a conventional AHE, which is proportional to $M$, but is in accordance with the context of the topological Hall effect (THE) characteristic of the "non-coplanar" magnetic structure with finite scalar spin chirality defined as $\chi_S = \boldsymbol{S_i} \cdot (\boldsymbol{S_j} \times \boldsymbol{S_k})$ (See inset of Fig. 5A) (*38*). Although non-coplanar spin textures are typically known to be appurtenant to magnets with non-centrosymmetric or frustrated crystal structures, they are also ubiquitous in

$M$-site of orthorhombic $LnMO_3$ including $NdMO_3$ ($M$: 3$d$ transition metal) (*40*). For example, in $NdMO_3$ ($M$ = Cr, Fe, etc.), employing Bertaut's notation (*41*), a non-coplanar $G_bA_aF_c$ spin texture of $M^{3+}$ moments has been widely confirmed. On the other hand, for $Nd^{3+}$ moments, $G_bA_a$ ($M$ = Sc, In) or $C_c$ ($M$ = Cr, Co, Ga) textures are commonly observed while both are coplanar (*42–46*). Considering the more itinerant nature of Ru-4$d$ electron, realization of the non-coplanar $G_bA_aF_c$ spin structure of $Ru^{3+}$ moments is unlikely in NRO, thus we speculate that finite $\chi_S$ is mediated by the spin texture of $Nd^{3+}$ moments. We are aware that this speculation should be supported by direct observation of magnetic structure or measurements of elemental-sensitive techniques such as neutron diffraction and x-ray magnetic circular dichroism. However, because of the thin film nature of the sample and the low ordering temperature (~1 K), they are not feasible at present. Therefore, we propose a possible mechanism based on the crystal symmetry as discussed below (Also see Supplementary Texts 6 and 7).

As a starting point, we hypothesize that $Nd^{3+}$ moments in NRO have $G_bA_a$ order as the ground state (Fig. S8C) and turn into ferromagnetic (FM) order (Fig. S8E) at ~2.5 T, where THE disappears (Fig. 5A), since $\chi_S$ should vanish in the FM state. This type of transition is indeed reported in $GdFeO_3$, where $Gd^{3+}$ moments take $G_bA_a$ ground state and turn into FM state at about 5 T and 2 K (*47*). As mentioned above, the spin structure of $G_bA_a$ order is coplanar, where $Nd^{3+}$ moments are confined within the $ab$-plane of orthorhombic setting. In our NRO film, the $ab$-plane is perpendicular to the film surface, which is confirmed by electron beam diffraction and RSM (See Figs. S3, S4, and Supplementary Text 3). Thus, when a magnetic field is applied perpendicular to the film surface, it is parallel to the $ab$-plane. With increasing the magnetic field, $Nd^{3+}$ moments are aligned toward the field direction, leading to an induced FM order. We suggest, during this transition, $Nd^{3+}$ moments point toward the $c$-axis and form a non-coplanar spin texture as presented in Fig. 5D (Also see Fig. S8D) instead of simply rotating in the $ab$-plane, which we conjecture can be a source of finite $\chi_S$. Itinerant Ru-4$d$ electrons interact with $Nd^{3+}$ moments and acquire $\chi_S$, resulting in the observed THE. Here in NRO, because of the square lattice, the definition of $\chi_S$ does not seem as straightforward as in the case of well-known triangular lattices. Yet, by summing up $\chi_S$ from all the possible combinations of the three spins, it can be achieved. Although a more detailed discussion is provided in Supplementary Text 7, $\chi_S$ from each contribution is indeed canceled out when the crystal structure has a high symmetry. We thus speculate that there must be some causes for breaking the bulk crystal symmetry. Although it is intriguing to elucidate the magnetic structure of NRO and authentic origin of the emergent THE, this is beyond the scope of this report and remains future work.

**Discussion**

In conclusion, we have successfully stabilized perovskite LRO and NRO with $Ru^{3+}$ in the epitaxial thin film form as a new materials platform to investigate the interaction between Ru-$4d$ and $Ln$-$4f$ electrons. Magnetotransport measurements highlight the clear difference between LRO and NRO reflecting the absence/presence of magnetic moments on $Ln$-site as we designed. Especially, only NRO exhibits an anomalous Hall effect, wherein interaction among $Ru^{3+}$ moments below 100 K, and between $Nd^{3+}$ and $Ru^{3+}$ moments below 20 K play important roles. Above all, we capture an unconventional AHE, namely topological Hall effect, below 1 K suggestive of the realization of a non-coplanar spin texture with finite scalar spin chirality. We propose a reasonable mechanism to realize the spin texture based on the common spin configuration of orthorhombic perovskite Nd$M$O$_3$, which will be clarified by further studies for direct evidence.

**Figures and Tables**

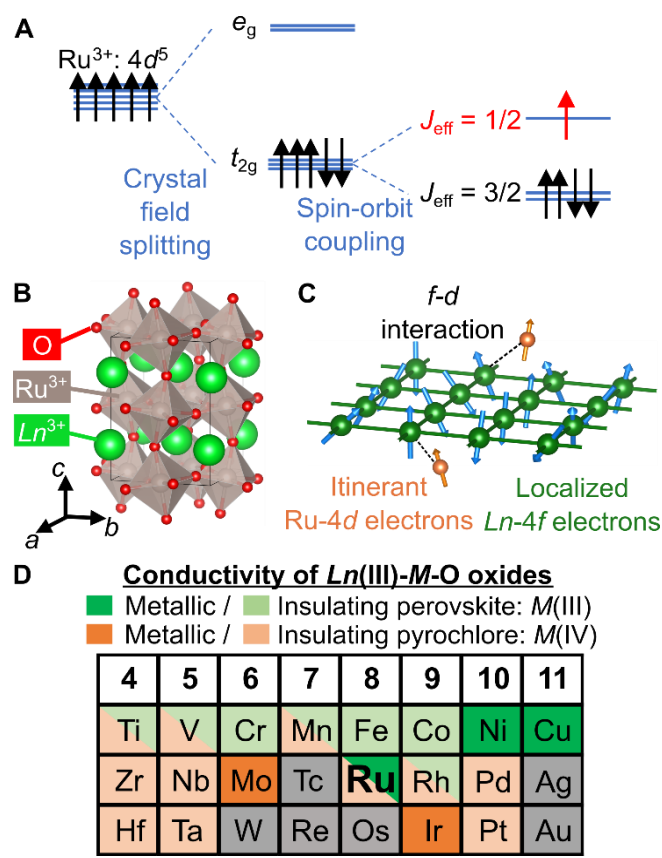

**Fig. 1. Concept of this study.** (**A**) Band splitting and $4d^5$ electron configuration of $Ru^{3+}$ ion. (**B**) Schematic of orthorhombic *Pbnm* perovskite structure illustrated by VESTA (*48*). (**C**) Schematic of the interaction between itinerant Ru-$4d$ and localized *Ln*-$4f$ moments with a non-coplanar ordering. (**D**) Summary of the electrical ground states of *Ln*(III)-*M*-O (*M*: transition metal) perovskite and pyrochlore oxides that have been reported (*16*, *26–29*, *49–55*). The gray color indicates no reports related to the elements.

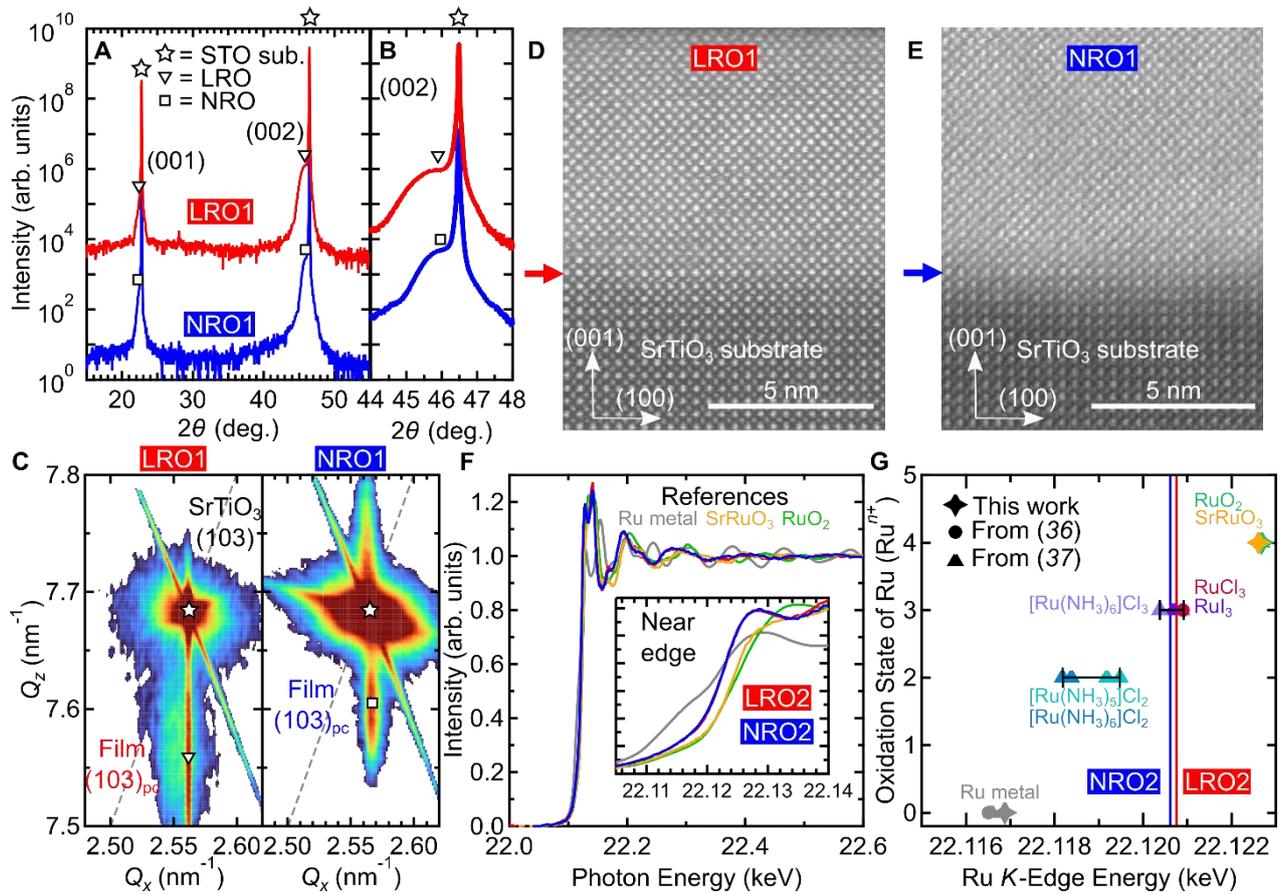

**Fig. 2. Structural properties and valence states of LaRuO$_3$ and NdRuO$_3$ thin films.** (**A**) X-ray diffraction (XRD) 2$\theta$-$\theta$ scan profiles and (**B**) magnified data around the (002) peaks of LaRuO$_3$ and NdRuO$_3$ thin films on SrTiO$_3$ (001) substrates. Peaks for the substrate, LaRuO$_3$, and NdRuO$_3$ are indicated by stars, triangles, and squares, respectively. (**C**) The XRD reciprocal space mappings around SrTiO$_3$ (103) peak. The gray broken lines indicate the relaxation lines where a fully relaxed cubic film peak is supposed to be located. TEM image of (**D**) LaRuO$_3$ and (**E**) NdRuO$_3$ thin films. The red and blue arrows indicate the position of the interface between the film and substrate. (**F**) Ru K-edge XAS spectra of LaRuO$_3$ and NdRuO$_3$ thin films in comparison with those of Ru$^0$ metal, Ru$^{4+}$O$_2$ and SrRu$^{4+}$O$_3$ as references. The intensity was normalized so that the averaged edge jump became unity. The inset shows the enlarged view of near-edge structures. (**G**) Oxidation state of Ru (Ru$^{n+}$) versus Ru K-edge energy, which is defined as the energy where the normalized intensity equals 0.5. The vertical red and blue lines correspond to the Ru K-edge energies of the LaRuO$_3$ and NdRuO$_3$ thin films, respectively. Symbols of stars, circles, and triangles refer to the data from present measurement, K. Fujiwara *et al.* (*36*), and SPring-8 BENTEN database (*37*), respectively.

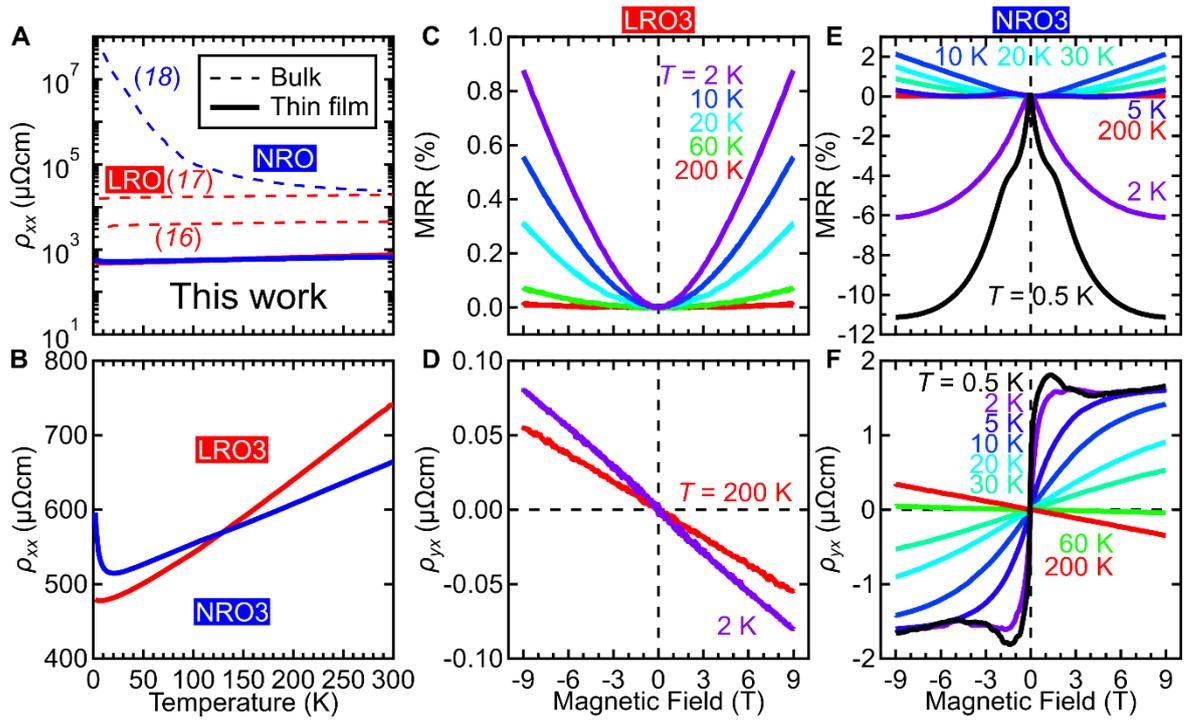

**Fig. 3. Magnetotransport properties of LaRuO$_3$ and NdRuO$_3$ thin films.** (**A**), (**B**) Temperature dependence of longitudinal resistivity ($\rho_{xx}$) compared with those in previous reports of polycrystalline bulk samples (*16–18*). Magnetic field dependence of magnetoresistance ratio (MRR) and Hall resistivity ($\rho_{yx}$) for (**C**), (**D**) LaRuO$_3$ and (**E**), (**F**) NdRuO$_3$ at selected temperatures.

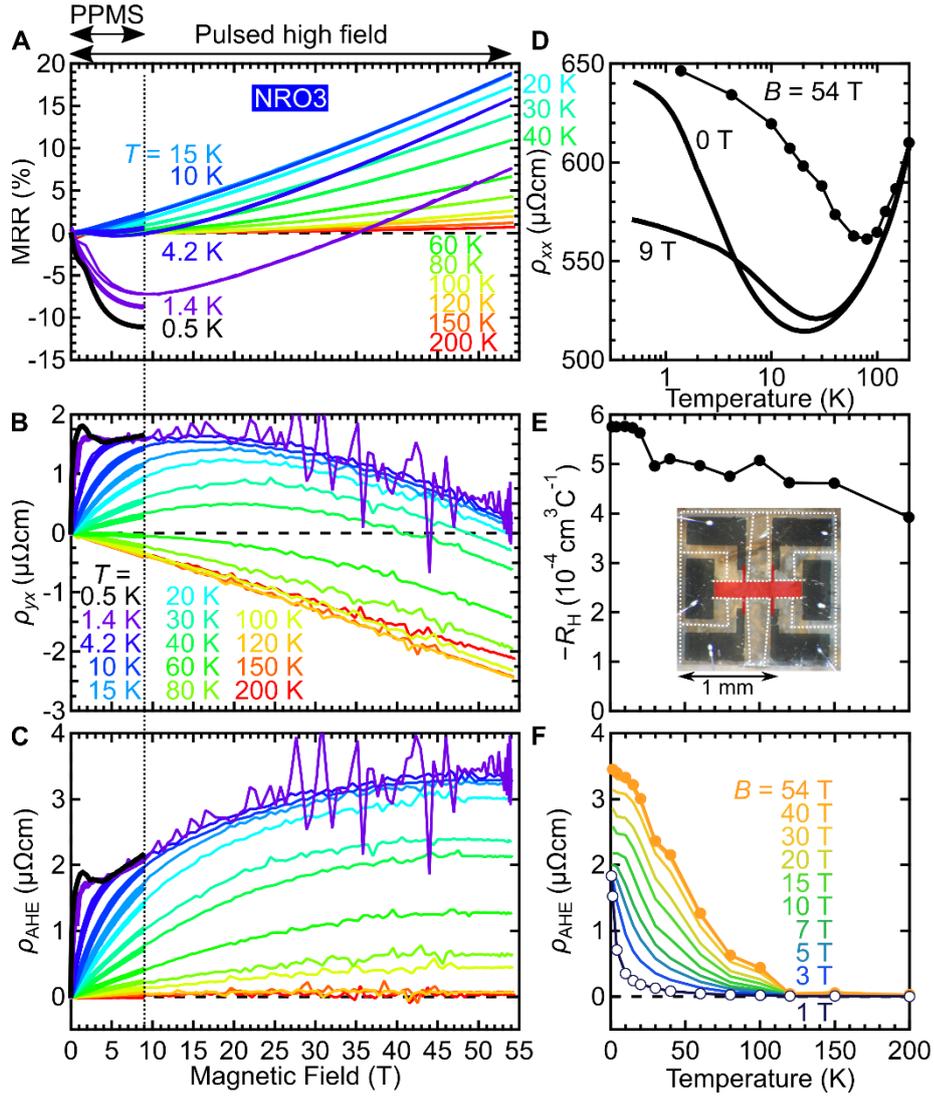

**Fig. 4. Magnetotransport properties of NdRuO$_3$ thin film measured with a pulsed high field magnet.** Magnetic field dependences of (**A**) MRR, (**B**) $\rho_{yx}$, and (**C**) anomalous Hall resistivity ($\rho_{AHE}$) at various temperatures. See the main text for the definition of $\rho_{AHE}$. Both of the results measured with a PPMS (thick lines below 9 T) and the pulsed high field magnet (thin lines) are shown. The vertical dotted line indicates the position of $B = 9$ T. The results at 1.4 K are noisy because of the low measurement current which is employed to suppress the Joule heating (See Materials and Methods for details). (**D**) Temperature dependence of $\rho_{xx}$ at $B = 0$, 9 (with a PPMS) and 54 T (with the pulsed high field magnet). (**E**) Temperature dependence of Hall coefficient ($R_H$) deduced by a linear fitting of $\rho_{yx}$ between 45 and 55 T. An optical image of the device structure for the transport measurements. Ni/Au electrodes (dark areas) are deposited on the film and then it is scribed (white dotted lines indicate the scribed lines) to fabricate a Hall bar structure highlighted by the red colored area. (**F**) Temperature dependence of $\rho_{AHE}$ at selected magnetic field. The curves for $B = 1$ and 54 T are presented by lines with open and filled circles.

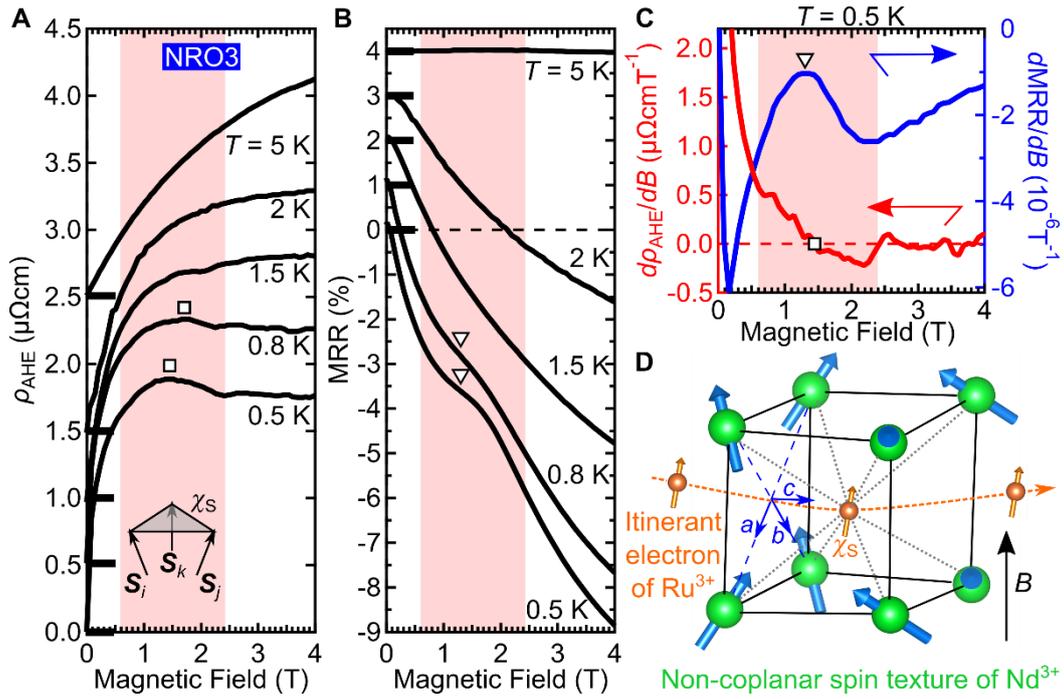

**Fig. 5. Unconventional Hall effect in NdRuO₃ thin film.** Magnetic field dependence of (**A**) $\rho_{AHE}$ and (**B**) MRR in the NRO film below 5 K at low-field region ($B \leq 4$ T) with vertical offsets for clarity. The results of 0.5, 2, and 5 K are identical to the ones presented in Figs. 3(**E**) and 3(**F**). For each curve in (**A**) and (**B**), its origin at $B = 0$ is indicated by a thick horizontal bar. Inset of (**A**) is a schematic of a non-coplanar spin texture that produces scalar spin chirality $\chi_S$. The peak position of AHE and inflection point of MRR are indicated by square and triangle, respectively. Magnetic field dependence of (**C**) magnetic field derivatives of $\rho_{AHE}$ (red curve, left axis) and MRR (blue curve, right axis) at 0.5 K. The square and triangle in (**C**) indicate the zero point in $d\rho_{AHE}/dB$ (i.e., peak in $\rho_{AHE}$) and the peak in dMRR/dB (i.e., an inflection point in MRR) around 1.3 T, respectively. The red-colored areas in (**A**)–(**C**) highlight the field region where the anomalies are observed at 0.5 K. (**D**) Schematic of a possible mechanism to generate the topological Hall effect. Itinerant electrons of $Ru^{3+}$ interact with localized $Nd^{3+}$ moments and acquire $\chi_S$ induced by non-coplanar spin texture of $Nd^{3+}$. Note that $Ru^{3+}$ locates at the body center of the deformed cube formed by $Nd^{3+}$.